
%
%
\nonstopmode
\input phyzzx.tex
\input tables.tex
\tolerance=5000
\sequentialequations
\overfullrule=0pt
\twelvepoint
\nopubblock
\font\anti=cmbsy10
\font\footrm=cmr9
\font\tenrm=cmr10
%
%
\line{\hfill IASSNS-HEP-92/78}
\line{\hfill CERN-TH-7105/93}
\line{November 1993\hfill }
\titlepage
\vskip1.0cm
\title{Microscopic formulation of the hierarchy of 
quantized Hall states}
\vskip0.5cm
\author{Martin Greiter\foot{Research
supported in parts by DOE grant DE-FG02-90ER40542}\foot{Present
address: Theory Division, CERN, CH 1211 Geneva 23;
greiter@cernvm.cern.ch}} 

\vskip.5cm
\centerline{{\it School of Natural Sciences}}
\centerline{{\it Institute for Advanced Study}}
\centerline{{\it Olden Lane}}
\centerline{{\it Princeton, N.J.\ 08540}}
\vskip.2cm
\abstract{
Explicit wave functions for the hierarchy of fractionally quantized Hall
states are proposed, and a method for integrating out the quasiparticle
coordinates in the spherical geometry is developed.  Their energies
and overlaps with the exact ground states for small numbers of particles
with Coulomb interactions are found to be excellent.
We then generalize the adiabatic transport argument of Arovas, Schrieffer,
and Wilczek to evaluate quasiparticle charges and statistics, and
show that none of the proposed states is the exact ground state of
any model Hamiltonian with two-body interactions only.
}
\endpage

\def\prl{{\sl Phys.\ Rev.\ Lett.\ }}

\def\prb{{\sl Phys.\ Rev.\ B\ }}

\def\npb{{\sl Nucl.\ Phys.\ B\ }}

\def\ss{{\sl Surface Science }}

\REF\laugha{R.B.~Laughlin,
\prl {\bf 50} (1983) 1395.}

\REF\num{F.D.M.~Haldane in R.E.~Prange and S.M.~Girvin (eds.),
{\sl The Quantum Hall Effect} (Springer, New York, 1990).}

\REF\halda{F.D.M.~Haldane,
\prl {\bf 51} (1983) 605.}

\REF\laughb{R.B.~Laughlin,
\ss {\bf 142} (1984) 163.}

\REF\halpa{B.I.~Halperin,
\prl {\bf 52} (1984) 1583.}

\REF\gwexact{
M. Greiter and F. Wilczek,
\npb {\bf 370} (1992) 577.}


\REF\yang{A similar proposal has been advocated independently by
J.~Yang, Houston preprint (1993);
I wish to thank B.I.~Halperin for communicating this work to me.}

\REF\asw{D.~Arovas, J.R.~Schrieffer, and F.~Wilczek,
\prl {\bf 53} (1984) 722.}

\REF\read{N.~Read, \prl {\bf 65} (1990) 1502;
B.~Blok and X.G.~Wen, \prb {\bf 41} (1990) 8145.}

\REF\halpb{B.I.~Halperin, {\it Helv.\ Phys.\ Acta}\ {\bf 56} (1983) 75.}

\REF\jain{J.K.~Jain, \prl {\bf 63} (1989) 199 and \prb {\bf 40} (1989) 8079;
J.K. Jain, S.A. Kivelson, and N. Trivedi, \prl {\bf 64} (1990) 1297.}

\REF\groot{M.~Greiter, {\it Root configurations and many body interactions
for fractionally quantized Hall states}, Institute for Advanced Study
report IASSNS-HEP-92/77.}

\REF\gmcd{M.~Greiter and I.A.~McDonald, {\it Hierarchy of quantized Hall
states in double layer electron systems},
to appear in \npb [FS].}

\REF\gww{M.~Greiter, X.G.~Wen, and F.~Wilczek,
\prl {\bf 66} (1991) 3205 and \npb {\bf 374} (1992) 567.}

\chapter{Introduction}

Most of our understanding of the
fractionally quantized Hall effect is based
on a highly original trial wave function [\laugha ]
for the ground state at filling fractions $1/m$,
where $m$ is an odd integer.
It describes
an incompressible quantum fluid with fractionally charged
excitations,
and has been shown [\num ]
to capture the correct universality classes
at those fillings.
The hierarchy \hbox{[\halda -\halpa ]}
is an extension of this picture to other
rational filling fractions, based on the idea that
the quasiparticle excitations themselves condense
into a Laughlin-Jastrow type fluid.
This procedure can be iterated, and
all odd denominator filling fractions appear within this picture,
at various levels of the hierarchy.

The problem with the existing formulations of the hierarchy
is that they only predict the corresponding
universality classes, specified by certain quantum
numbers, but do not provide us with explicit wave functions.
It is the aim of the present letter to fill this gap.
In the following chapter, we will propose such wave functions,
develop a method for integrating out the quasiparticle
coordinates, and present the results of extensive numerical
studies, which provide strong evidence in support of our
proposal.
In the chapter thereafter, we obtain general formulas for the
quasiparticle charges and statistics, and present preliminary
results for energy gaps.  We then present a simple proof
that none of the proposed hierarchy states is the exact ground
of any model Hamiltonian containing two body interactions only,
and conclude with 
comments on Jain's construction, double layer electron systems and
paired Hall states.

\chapter{Microscopic Formulation}

The general principles are most easily demonstrated by considering
first a specific example.  The simplest hierarchy state we can
write down is a $m$ parent state with a $p$ daughter of quasi-holes, denoted
by $[m,+p]$, where the $+$ sign indicates that we have quasi-holes
rather then quasi-electrons.  For convenience, we work
in the spherical geometry [\halda ].
The wave function we wish to propose
for this state is
$$
\eqalign{
\Psi_{[m,+p]}[u,v] ~=~ \int D[a,b]\
&\prod_{k<l}^{N_1}\, (\bar a_k \bar b_l - \bar a_l \bar b_k)^p\,\times
\cr\noalign{\vskip 4pt}
\times\,
&\prod_{k=1}^{N_1}\,\prod_{i=1}^N\, (a_k u_i - b_k v_i)
\ \prod_{i<j}^N\, (u_i v_j - u_j v_i)^m
\cr}
\eqn\mphole
$$
where $(u_i,v_i)$ with $i=1,...\,N$
are the spinor coordinates of the electrons
on the sphere,
$(a_k,b_k)$ with $k=1,...\,N_1$
are those 
of the quasiparticles,
and $(\bar a_k,\bar b_k)$ their complex conjugates.
The interpretation of \mphole\ is straight forward:
we take a Laughlin $1/m$-state for $N$ electrons,
create $N_1$ quasi-holes, weight the different quasi-hole
configurations by a Jastrow factor taken to an even power $p$,
and integrate the quasi-hole coordinates over the unit
sphere.

To obtain a $[m,-p]$ state, we just need to replace the quasi-holes
in \mphole\ by quasi-electrons, that is,
$$
\prod_{k=1}^{N_1}\,\prod_{i=1}^N\, (a_k u_i - b_k v_i)
\rightarrow
\prod_{k=1}^{N_1}\,\prod_{i=1}^N\,
(\bar b_k {\partial \over\partial u_i} -
\bar a_k {\partial \over\partial v_i}).
\eqn\phchange
$$

This wave function would, of course, not be of much practical use if
we had no method to perform the quasiparticle integration.
Fortunately, this integration is rather straight forward.
Expressing the spinor coordinates for a given particle in terms
of the polar and
azimuthal angles $\phi $ and $\theta $ on the sphere,
$a=\cos\! {\theta\over 2}\, e^{+{i\over 2}\phi}$ and
$b=\sin\! {\theta\over 2}\, e^{-{i\over 2}\phi}$, and substituting
$\int\! D(a,b) \equiv {1\over 4\pi}\! \int\! d\Omega$, we find
$$
\int\! D(a,b)\, \bar a^{n'}\, \bar b^{\,2S'-n'} a^n\, b^{\,2S-n}\,
=\, {n!\, (2S-n)! \over (2S+1)!}\, \delta_{nn'}\, \delta_{2S,2S'}.
\eqn\method
$$
In any given polynomials in the present context, the integer $2S$
is given by the number of flux quanta through the sphere; it is the same
for all terms in an expansion, and we may absorb the $(2S+1)!$ on
the right of \method\ in the overall normalization of the wave function.
Thus we may,
instead of performing the integration, replace
$\bar a \rightarrow \partial_a$,
$\bar b \rightarrow \partial_b$
and then take $a,b \rightarrow 0$.  This leaves us with
a large polynomial with derivatives, which is in principle
not significantly more difficult to expand then say
a Laughlin state.

For our wave function to be non-zero the
total degree $2S$ of the polynomials in $(\bar a_k,\bar b_k)$
and $(a_k, b_k)$ has to be the same; this leads to the constraint
$$
p\, (N_1 - 1)\,=\,N.
$$
The total number of Dirac flux quanta seen by the physical electrons is
therefore
$$
\eqalignno{
2S\, &=\, m\,(N-1) \pm N_1 \cr\noalign{\vskip 4pt}
     &=\,\left( m\pm {1\over p}\right) N - m \pm 1\, ,\cr}
$$
where the $+$ sign corresponds to quasi-holes, and the $-$ sign to
quasi-electrons.
The inverse filling fraction and the flux shift on
the sphere, defined via
$$
2S\,=\,{1\over\nu }\, N - N_{\hbox{\footrm shift}} \,,
\eqn\fnrel
$$
are thus as predicted by Haldane [\halda ], and consisted with the result
of exact diagonalization studies for small numbers of particles.

Before proceeding to the general case, we wish to
refine our construction.  If we take
a close look at the trial wave function involving
quasi-electrons, we note that we cannot obtain a $[5,-2]$ state
($\nu =2/9$)
from a $[3,-2]$ state ($\nu =2/5$) by
just multiplying the latter with a Jastrow
factor squared, as we would expect from a very general argument
based on an adiabatic localization of magnetic flux [\gwexact ].  To
circumvent this problem, we are led to consider an alternative trial
wave function for the quasi-electron,
$$
\Psi_m^{+}[u,v] ~=~
\Bigl[\,\,\prod_{i=1}^N\,
(\bar b \,{\partial \over\partial u_i} -
 \bar a \,{\partial \over\partial v_i})
\, \prod_{i<j}^N\, (u_i v_j - u_j v_i)^2\,\Bigr]
\,\, \prod_{i<j}^N\, (u_i v_j - u_j v_i)^{m-2},
\eqn\gquasi
$$
where the derivatives act only on the term in the square brackets.
This trial wave function
turns out to be energetically favorable over Laughlins
---which is not surprising, as it preserves the maximal
possible number of Jastrow factors.  Indeed, when calculating
the overlap between a $2/5$ state and the exact ground state for
Coulomb interactions for \hbox{$N=6$}, we find 0.9985 if we use Laughlin's
quasi-particle and 0.9995 if we use \gquasi .

Extending our construction to hierarchy wave
functions in general, we write the recursion relations
$$
\eqalign{
\Psi_{[m,+p_1,...\alpha_n p_n]}[u,v] &=
\Psi_{[p_1,...\alpha_n p_n]}[\partial_a,\partial_b]
\prod_{k=1}^{N_1}\prod_{i=1}^N (a_k u_i - b_k v_i)
\prod_{i<j}^N (u_i v_j - u_j v_i)^m
\cr\noalign{\vskip5pt}
\Psi_{[m,-p_1,...\alpha_n p_n]}[u,v] &=
\prod_{i<j}^N (u_i v_j - u_j v_i)^{m-2} \,\times
\cr\noalign{\vskip 3pt}
&\times \Psi_{[p_1,...\alpha_n p_n]}[\partial_{\bar a},\partial_{\bar b}]
\prod_{k=1}^{N_1}\prod_{i=1}^N
({\bar b}_k \partial_{u_i} - {\bar a}_k \partial_{v_i})
\prod_{i<j}^N (u_i v_j - u_j v_i)^2 .\cr}
\eqn\gen
$$
Note that the quasiparticle wave function
$\Psi_{[p_1,...\alpha_n p_n]}[a,b]$
is always symmetric, as all the $p$'s are even;
$\Psi_{[m,\alpha_1 p_1,...\alpha_n p_n]}[u,v]$ is symmetric if
$m$ is even, or antisymmetric if $m$ is odd, as appropriate for an
electron wave function [\yang ].

These wave functions lead to the iterative equations
$$
\eqalign{
{1 \over \nu} \equiv [m,\alpha_{1} p_{1}, ...,\alpha_{n} p_{n}]
&= m + {{\alpha_{1}} \over [p_1, \alpha_2 p_2,..., \alpha_n p_n]}
\cr\noalign{\vskip 4pt}
N_{\hbox{\footrm shift}} \equiv \{ m, \alpha_1 p_1,..., \alpha_n p_n \}
&= m - \alpha_1 {{\{p_1, \alpha_2 p_2,..., \alpha_n p_n \}}
\over {[p_1, \alpha_2 p_2,..., \alpha_n p_n]}} \cr}
\eqn\iter
$$
for inverse filling fractions and flux shifts on the sphere, and thus
reproduce the universality classes predicted by Haldane [\halda ].

For some of
the most important hierarchy states, we have numerically evaluated
overlaps with the exact ground state and energy expectation values
for small numbers of particles with
Coulomb interactions.  The results confirm the validity of our
trial wave functions; they are shown in Table 1.  Note in particular,
that the hierarchy states at $\nu =2/5$ and $\nu =3/7$
rate considerably better
then Laughlin's original $1/3$ state, which has been included for
comparison.
It should also be noted that a hierarchy $2/3$ state is
energetically less favorable then
the particle hole conjugate of a Laughlin $1/3$ state, while
a hierarchy $2/7$ state is better then a $2/7$ state
constructed by multiplying this particle hole conjugate
by a Jastrow factor squared.

\chapter{Quasiparticle charge and statistics}

The elementary excitations of the general hierarchy states \gen\ are just
conventional quasiparticles (quasi-holes or quasi-electrons) in the
last daughter fluid $\Psi_{[p_n]} $.  To determine their
charge, we just have to observe how the flux--particle number relationship
%
%
is altered by the presence of a quasiparticle; this yields
$$
\eqalign{
e^* ~&\equiv ~(m,\alpha_{1} p_{1},...,\alpha_{n} p_{n})
\over [m,\alpha_{1} p_{1},...,\alpha_{n} p_{n}]}~=
\cr\noalign{\vskip 4pt}
&=~ {(-1)^n\,\alpha_1\cdot\alpha_2\cdot ...\cdot\alpha_n
\over \hbox{$\, $denominator of filling fraction $\nu $}}\, ,\cr}
\eqn\charge
$$
where a negative sign is meant to indicate that a quasi-hole
in the last daughter fluid corresponds to a quasi-electron
in the electron fluid and vice versa.  For example, the
charge of a quasiparticle in a $2/5$ state is $e^*=1/5$.

Turning to the quasiparticle statistics, let us first recall
the adiabatic transport argument of Arovas, Schrieffer, and Wilczek [\asw ]
for Laughlin-Jastrow type states.
The quasiparticle charge
$e^{*}$ is obtained by equating the geometric
phase $\gamma$ acquired by the wave function
as a quasi-hole is adiabatically carried around a closed loop
with the corresponding Aharonov-Bohm phase,
$$
i\oint \bra{\psi_m^-}{d\over dt}\ket{\psi_m^-} ~=~ -2\pi e^* \phi ,
$$
where $\phi$ is the flux through the loop measured in Dirac quanta.
Evaluation of the geometric phase yields $\gamma =-2\pi\VEV{N}$,
where $\VEV{N}$ is the average number of particles inside the loop.
Thus $e^*=1/m$.  If there is a second quasi-hole in
the loop, there will be a deficit of $1/m$ in $\VEV{N}$, and we obtain
an additional phase $\Delta \gamma={2\pi/m}$.  This additional phase
is identified as quasiparticle statistics $\theta =1/m$.

This argument can be generalized to the hierarchy; the only difference
is that the geometric phase depends now on
the average number
$\VEV{N_n}$ of those quasiparticles in the
loop which condense as the last daughter fluid,
$$
\gamma=-2\pi (-1)^n\,\alpha_1\cdot\alpha_2\cdot ...\cdot\alpha_n\, \VEV{N_n}.
$$
The statistics depends thus on how
$\VEV{N_n}$ is altered by the presence of a quasiparticle excitation
in the loop;
that is to say, on the amount we have to shift $N_n$ on the sphere to
accommodate a quasiparticle while keeping $2S$ fixed.  This yields
$$
\theta  ~=~
{(m,\alpha_{1} p_{1},...,\alpha_{n} p_{n})
\over  |(m,\alpha_{1} p_{1},
...,\alpha_{n-1} p_{n-1})|}\,,
\eqn\stat
$$
where $(m,\alpha_{1} p_{1}...)$ are the quasiparticle charges given by
\charge .  This result is consistent
with Halperin's predictions [\halpa ] and
with conclusions drawn from the effective field theory [\read ].
To give an example, the mutual statistics of the elementary
quasiparticle excitations in a $2/5$ state is $\theta =3/5$.  This
might look implausible at first sight.  To make it plausible,
let us consider pairs of these quasiparticles.
They will have charge $2/5$ and statistics
$4\times 3/5=12/5\rightarrow 2/5$, since each particle
in a pair will pick up a statistical phase from each of the particles in
any other pair.  The statistics of this excitation is thus consistent with
the statistics of the charge $2/5$ quasiparticle we would obtain by
directly inserting a Laughlin quasiparticle in the parent fluid.  This
consistency, which carries through for quasiparticles constructed by
inserting flux tubes at any level of a general hierarchy state, is most
essential to the consistency of the whole theory. For otherwise we could
combine quasi-electrons and quasi-holes at different levels of the hierarchy
to form new excitations which carry statistics but no charge, and
our charged quasiparticles might no longer be the elementary excitations.

The explicit wave functions proposed above can further be used
to calculate quasiparticle energies, and in particular
energy gaps.  One way to estimate the gap for systems with
small numbers of particles is to insert a quasi-electron at the north
and a quasi-hole at the south pole of the sphere; for our $2/5$ state,
we find energy gaps of 0.07569 and 0.06970 for $N=6$ and $8$, respectively,
which ought to be compared with 0.07505 and 0.6809 obtained by exact
diagonalization of the corresponding Hamiltonians.


\chapter{Model Hamiltonian}

The question we wish to address next is whether we can construct
a model Hamiltonian involving two body potentials only which singles
out any of our hierarchy states as unique and exact ground state.
Let us assume such a Hamiltonian would exist, and write $H \ket{\psi }=0$.
In a finite geometry, any two body interaction between
particles in the lowest Landau level can be parametrized by a finite
number of pseudopotentials $v_l$, which just denote the potential
energy cost of having relative angular momenta $l$ between two
particles.  Thus we may write
$$
H\ket{\psi }~=~
\bigl( \sum_{l=0}^{2S} v_{l} H_{l}\bigr) \ket{\psi }~=~
\sum_{l=0}^{2S} v_{l}\,(H_{l}\ket{\psi })~=~0.
\eqn\ham
$$
For the explicit wave functions proposed above we
find numerically that there is only one nontrivial solution to \ham :
$$
v_l=(2S - l)(2S - l+ 1) + \hbox{const.}
\eqn\pseudo
$$
This solution, however, corresponds to a Hamiltonian equal to the square of
the total angular momentum of the many body wave function
on the sphere,
\hbox{$H={\vec L}_{\hbox{\footrm tot}}^2$,} and can therefore not be used
to distinguish one trial wave function from another.
Thus we have shown that there is no model Hamiltonian involving two body
potentials only which singles out any of our hierarchy states as
the unique ground state.  The same is true for the trial wave functions
proposed by Halperin [\halpb ] and Jain [\jain ].

In this context it is perhaps propitious to note that interesting new
possibilities arise as one considers many body interaction
potentials [\groot ].
For example, one can identify the exact ground state of a model
Hamiltonian which demands that the wave function vanishes as the sixth
power as three particles approach each other; this state has filling
fraction $2/5$.

\chapter{Comments}

We wish to conclude with a few comments:

\pointbegin
Comparing ground state energies and overlaps of our hierarchy wave functions
with the trial wave functions proposed by Jain [\jain ],
we find that Jain's $2/5$
state is slightly better than ours, while our $3/7$ state is better than
Jain's.  We conjecture that the reason for this is as follows:  The
improved trial wave function for the quasi-electron \gquasi\ is still not
the best in town; the pair wave function proposed by Halperin [\halpb ]
and the trial wave function proposed by Jain [\jain ] are even better.
In fact, we conjecture that these two are identical, as indicated by
expansions of the wave functions for $\nu = 1/3$ on the sphere up to $N=6$.
Now suppose we wish to construct a hierarchy $2/5$ state
along the lines of \gen , but with
Jain's trial wave function for the quasi-electron instead of \gquasi .
For a $2/5$ state, we need half as many quasiparticles as there are
electrons in the liquid --- but this is to say,
the starting point for this state
is two filled Landau levels!  In this context, it is
of course not clear what one could
mean by a condensation of the quasi-particles
into a $p=2$ daughter fluid, as they are already condensed
in the sense that they occupy
all the states in the second Landau level.  It is clear, however,
that the trial wave function one obtains in the end will be identical
to the one proposed by Jain.  We are thus led to conjecture that a Jain
$2/5$ state is identical to
a hierarchy state constructed with Halperin's
pair wave function for the quasi-electron.
This identification
would not apply to higher levels of the hierarchy,
and would thus be
consistent with the fact that our $3/7$ state is slightly
better than Jain's.
\point
The generalization of our wave functions to the hierarchy of double layer
electron systems [\gmcd ] follows without incident.
\point
The methods developed above can be used to construct a trial wave functions
for other universality classes as well, most noteworthy among them
paired Hall states [\halpb ,\gww ]:
$$
\eqalign{
\Psi_{\hbox{\footrm phs}}^m[u,v] &= \,\,\hbox{\anti A} \int\! D[a,b]\,
\prod_{i<j}^n\, (a_i b_j - a_i b_j)^m\,\times
\cr\noalign{\vskip 4pt}
\times\,
&\prod_{i=1}^n \left( (u_i \bar a_i - v_i \bar b_i)
(u_{i+n} \bar a_i - v_{i+n} \bar b_i)\right)^{\hbox{$m(n-1)\over 2$}}\,
\prod_{i=1}^n (u_i v_{i+n} - u_{i+n} v_i)\,,
\cr}
\eqn\phs
$$
where \hbox{\anti A} denotes antisymmetrization,
$n$ is the number of electron pairs with center-of-mass coordinates
$(a_i,b_i)$, and $m$ is an even integer related to the filling
fraction via $\nu =4/m$.
\par

{\it Acknowledgements.}---
Much of my knowledge of the subject was acquired in numerous
discussions with F.D.M.\ Haldane, B.I.\ Halperin,
X.G.\ Wen and F.\ Wilczek; my deepest gratitude goes to all of them.
I further wish to thank I.A.\ McDonald for his critical reading
of the manuscript, and the Laboratoires de Physique Statistique
and Theorique at the Ecole Normale Superieure in Paris for hospitality.

\vfill\endpage

{\narrower{
TABLE 1: Overlaps and ground state energies per particle for some of the
most important hierarchy states.}
}
\bigskip\vskip8pt

\begintable
$\nu $ | $[m,\alpha_1 p_1...]$ | $N$ | $2S\!+\!1$ |
$\braket{\Psi_{\hbox{\footrm trial}}}{\Psi_{\hbox{\footrm ex}}}$ |
$E_{\hbox{\footrm trial}}$ | $E_{\hbox{\footrm ex}}$ &\cr
1/3 | $[3]$       | 6 | 16 | 0.9964 | -0.44995 | -0.45017 &\nr
2/5 | $[3,-2]$    | 6 | 12 | 0.9995 | -0.50036 | -0.50040 &\nr
    |             | 8 | 17 | 0.9988 | -0.48017 | -0.48024 &\nr
3/7 | $[3,-2,-2]$ | 6 | 10 | 0.9981 | -0.53950 | -0.53964 &\nr
2/3 | $[1,+2]$    | 8 | 13 | 0.9939 | -0.53372 | -0.53415 &\nr
    | -----       |   |    | 0.9990$^*$ | -0.53408$^*$ | &\nr
2/7 | $[3,+2]$    | 6 | 20 | 0.9906 | -0.40415 | -0.40436 &\nr
    | -----       |   |    | 0.9856$^\dagger $ | -0.40405$^\dagger$ | &\nr
2/9 | $[5,-2]$    | 6 | 22 | 0.9915 | -0.38695 | -0.38714 & 
\endtable
\vskip-3pt
\line{\hskip1.8truecm\tenrm
$*$ particle hole conjugate of Laughlin 1/3 state\hfill }
\vskip-6pt
\line{\hskip1.8truecm\tenrm
$\dagger$ particle hole conjugate of Laughlin 1/3 times Jastrow
factor squared\hfill }

\vfill
\endpage
\refout
\end